\documentclass[%
reprint,
 amsmath,amssymb,
 aps,
]{revtex4-2}

\usepackage{graphicx}
\usepackage{dcolumn}
\usepackage{bm}

\usepackage[caption=false]{subfig}
\usepackage{siunitx, soul, xcolor}
\usepackage{adjustbox}
\usepackage{gensymb}
\usepackage{bm}

\usepackage{booktabs} 

\usepackage{xparse,xcoffins}

\ExplSyntaxOn
\NewCoffin\imagecoffin
\NewCoffin\labelcoffin

\keys_define:nn { miguel/label }
 {
  label   .tl_set:N = \l_miguel_label_tl,
  labelbox .bool_set:N = \l_miguel_label_box_bool,
  labelbox .default:n = true,
  fontsize .tl_set:N = \l_miguel_label_size_tl,
  fontsize .initial:n = \footnotesize,
  pos .choice:,
  pos/nw .code:n = \tl_set:Nn \l_miguel_label_pos_tl { left,up },
  pos/ne .code:n = \tl_set:Nn \l_miguel_label_pos_tl { right,up },
  pos/sw .code:n = \tl_set:Nn \l_miguel_label_pos_tl { left,down },
  pos/se .code:n = \tl_set:Nn \l_miguel_label_pos_tl { right,down },
  pos/n .code:n = \tl_set:Nn \l_miguel_label_pos_tl { hc,up },
  pos/w .code:n = \tl_set:Nn \l_miguel_label_pos_tl { left,vc },
  pos/s .code:n = \tl_set:Nn \l_miguel_label_pos_tl { hc,down },
  pos/e .code:n = \tl_set:Nn \l_miguel_label_pos_tl { right,vc },
  pos .initial:n = nw,
  unknown .code:n   = \clist_put_right:Nx \l_miguel_label_clist
                       { \l_keys_key_tl = \exp_not:n { #1 } }
 }
\clist_new:N \l_miguel_label_clist
\box_new:N \l_miguel_label_box
\box_new:N \l_miguel_label_image_box

\NewDocumentCommand{\xincludegraphics}{O{}m}
 {
  \group_begin:
  \tl_clear:N \l_miguel_label_tl
  \clist_clear:N \l_miguel_label_clist
  \keys_set:nn { miguel/label } { #1 }
  \tl_if_empty:NTF \l_miguel_label_tl
   {
    \miguel_includegraphics:Vn \l_miguel_label_clist { #2 }
   }
   {
    \SetHorizontalCoffin\imagecoffin
     {
      \miguel_includegraphics:Vn \l_miguel_label_clist { #2 }
     }
    \SetHorizontalCoffin\labelcoffin
     {
      \raisebox{\depth}
       {
        \bool_if:NTF \l_miguel_label_box_bool
         { \fcolorbox{white}{white}{\l_miguel_label_size_tl\l_miguel_label_tl} }
         { \l_miguel_label_size_tl\l_miguel_label_tl }
       }
     }
    \SetVerticalPole\imagecoffin{left}{-0pt+\CoffinWidth\labelcoffin/2}
    \SetVerticalPole\imagecoffin{right}{\Width-3pt-\CoffinWidth\labelcoffin/2}
    \SetHorizontalPole\imagecoffin{up}{\Height+10pt-\CoffinHeight\labelcoffin/2}
    \SetHorizontalPole\imagecoffin{down}{3pt+\CoffinHeight\labelcoffin/2}
    \use:x{\JoinCoffins\imagecoffin[\l_miguel_label_pos_tl]\labelcoffin[vc,hc]} 
    \TypesetCoffin\imagecoffin
   }
   \group_end:
 }
\NewDocumentCommand{\setlabel}{m}
 {
  \keys_set:nn { miguel/label } { #1 }
 }

\cs_new_protected:Nn \miguel_includegraphics:nn
 {
  \includegraphics[#1]{#2}
 }
\cs_generate_variant:Nn \miguel_includegraphics:nn { V }

\ExplSyntaxOff

\begin{document}
\preprint{APS/123-QED}

\title{Significant ordinary Nernst effect contribution to spin-orbit torque harmonic Hall measurements in metallic structures}

\author{Igor Lyalin}
\email{igor.lyalin.4@gmail.com}
\affiliation{Department of Physics and Astronomy, Johns Hopkins University, Baltimore, Maryland 21218, USA}
\author{C. C. Chiang}
\affiliation{Department of Physics and Astronomy, Johns Hopkins University, Baltimore, Maryland 21218, USA}
\author{C. L. Chien}
\email{clchien@jhu.edu}
\affiliation{Department of Physics and Astronomy, Johns Hopkins University, Baltimore, Maryland 21218, USA}

\begin{abstract}
The harmonic Hall measurements are commonly used to quantify the spin-orbit torque in ferromagnet/normal metal bilayers. 
The contribution from ordinary Nernst effect to the second harmonic voltages is usually assumed to be negligible and omitted in the data analysis.
We show that in Cr/Co bilayers the ordinary Nernst effect cannot be neglected and can lead to
largely exaggerated values of spin-orbit torque efficiency.
Conducting additional experiments, we estimate the Nernst coefficient of Cr and find it comparable to the values reported for other metals.
Consequently, the ordinary Nernst effect should be carefully considered when employing the harmonic Hall technique to quantify spin-orbit torque in metals, while the results of some previous works might need to be re-examined.
\end{abstract}

\flushbottom
\maketitle
\thispagestyle{empty}

\section{Introduction}

Spin-orbit torque (SOT) is a promising energy-efficient way to manipulate magnetic order in bilayers of a ferromagnet (FM) and a nonmagnetic metal (NM)~\cite{liu_spin-torque_2011,miron_perpendicular_2011,liu_spin-torque_2012,manchon_current-induced_2019}.
Different SOT measurement techniques~\cite{pi_tilting_2010,liu_spin-torque_2011,emori_spin_2014,fan_quantifying_2014,pai_determination_2016} have been used to study the underlying spin Hall and spin Rashba-Edelstein effects~\cite{dyakonov_possibility_1971,hirsch_spin_1999,kato_observation_2004,rashba_1960,edelstein_spin_1990,sinova_universal_2004},
as well as to find material systems with large SOTs and optimize them for spintronic applications.
The harmonic Hall (HH) voltage measurement is one of the most widely used techniques utilized for these purposes, due to its relative technical availability and capability to quantify both, damping-like and field-like, components of SOT.

The HH measurement can be applied to FM/NM bilayers with either out-of-plane magnetization~\cite{pi_tilting_2010,kim_layer_2013}, or with in-plane magnetization~\cite{avci_interplay_2014}.
Here, we focus on the latter case.
In this measurement configuration, the external magnetic field is rotated in the film plane while the angular dependence of the harmonic Hall voltage is recorded.
The analysis of the angular and magnetic field dependencies allows to determine the spin-orbit torque.
The main assumption of the HH analysis is that the thermal effects are independent of the magnetic field, which allows to separate thermal and SOT contributions~\cite{avci_interplay_2014}.
This assumption holds true if the contribution from the ordinary Nernst effect (ONE) is negligibly small.
Previous works have demonstrated that the ONE can play a major role if the NM layer is a semi-metal or topological insulator~\cite{roschewsky_spin-orbit_2019,chi_spin_2020}.
However, for the case of metallic bilayers the ordinary Nernst effect is considered to be negligible.

In this work, we study spin-orbit torque in Co/Cr bilayers using harmonic Hall technique. 
Surprisingly, we find that depending on the stacking sequence, Co/Cr vs Cr/Co, damping-like SOT efficiency differs by an order of magnitude and has the opposite sign.
Performing experiments on Cr alone, we observe a magnetic field-dependent second harmonic Hall voltage consistent with the ordinary Nernst effect.
We show that in combination with a shunting effect the ONE can give rise to spurious SOT efficiencies on the order of $10^4 - 10^5$\,$\Omega^{-1}$\,m$^{-1}$.
A recent report of the giant orbital torque in Cr-based multilayers~\cite{sala_giant_2022} can be a consequence of this spurious effect.
Furthermore, we estimate the ordinary Nernst coefficient in Cr and find that it is comparable to the values reported for other metals, such as Cu.
Therefore, not just in Cr-based systems, but for metallic bilayers in general, the assumption of a negligible ONE in the harmonic Hall measurements is invalid.
Additional experiments have to be conducted to exclude the ONE contribution and to estimate the SOT accurately.  
\begin{figure*}[ht!]
\subfloat{\xincludegraphics[height=0.25\textwidth,label=(a)]{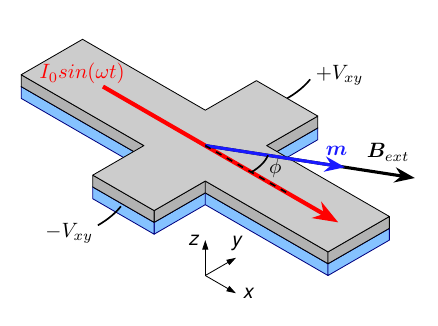}\label{fig:HH_schematics}}\hfill
\subfloat{\xincludegraphics[height=0.25\textwidth,label=(b)]{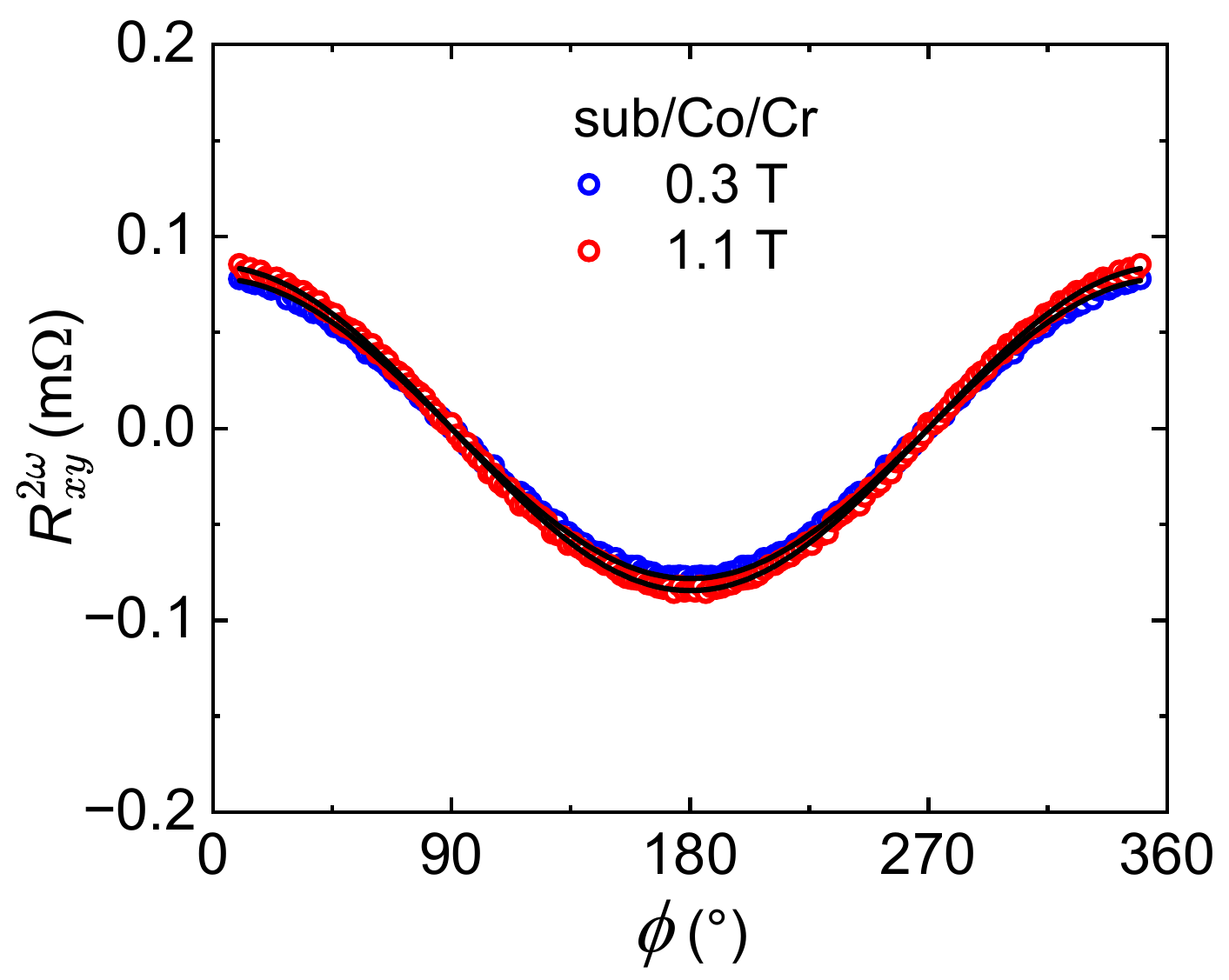}\label{fig:R2w_angular_scan_sub_Co_Cr}}\hspace{-0.0cm}
\subfloat{\xincludegraphics[height=0.25\textwidth,label=(c)]{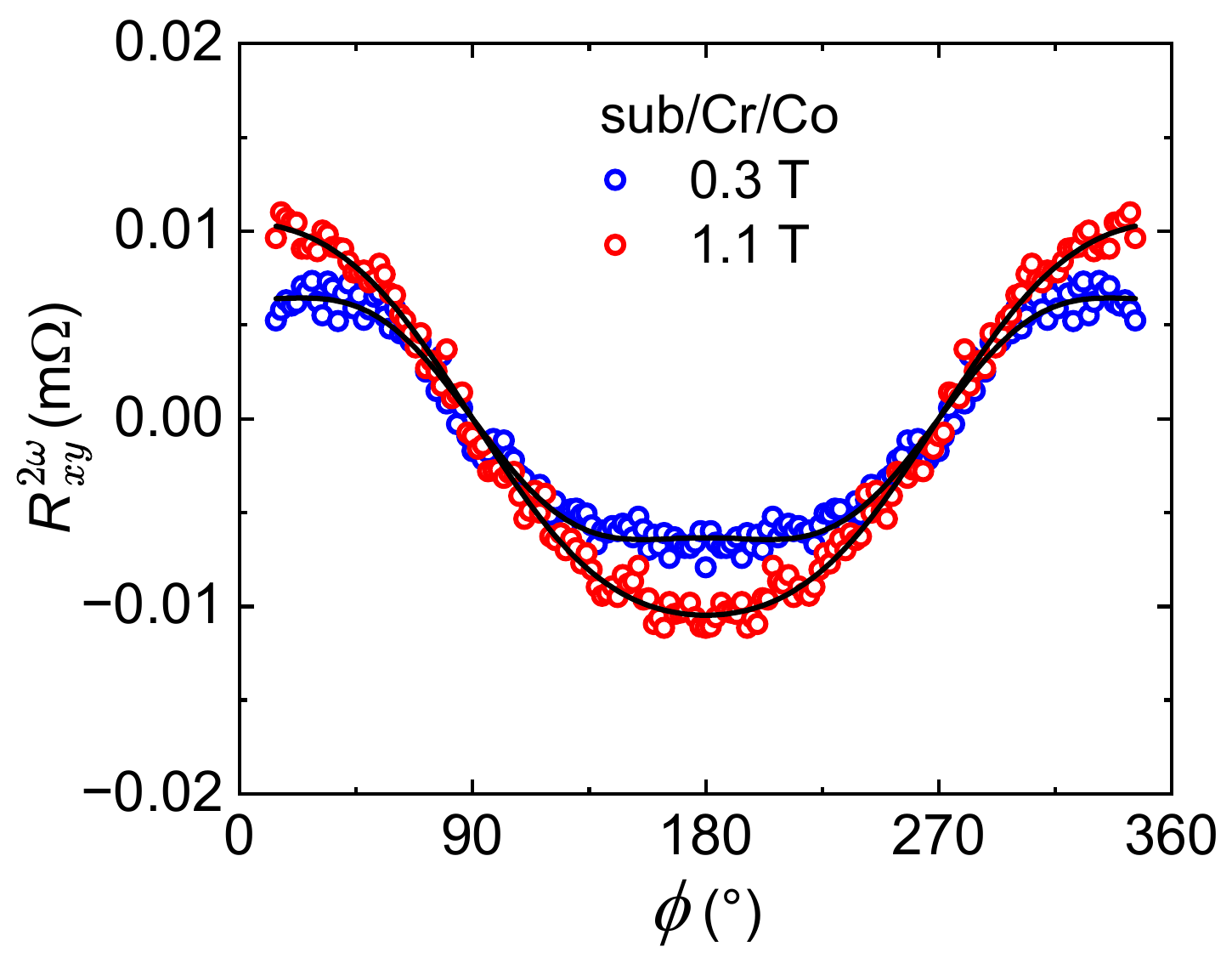}\label{fig:R2w_angular_scan_sub_Cr_Co}}
\vspace{-0.7cm}
\subfloat{\xincludegraphics[height=0.25\textwidth,label=(d)]{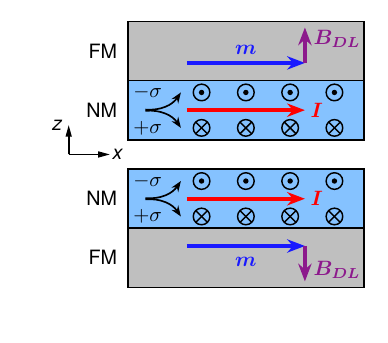}\label{fig:NM_FM_vs_FM_NM_schematics}}\hfill
\subfloat{\xincludegraphics[height=0.24\textwidth,label=(e)]{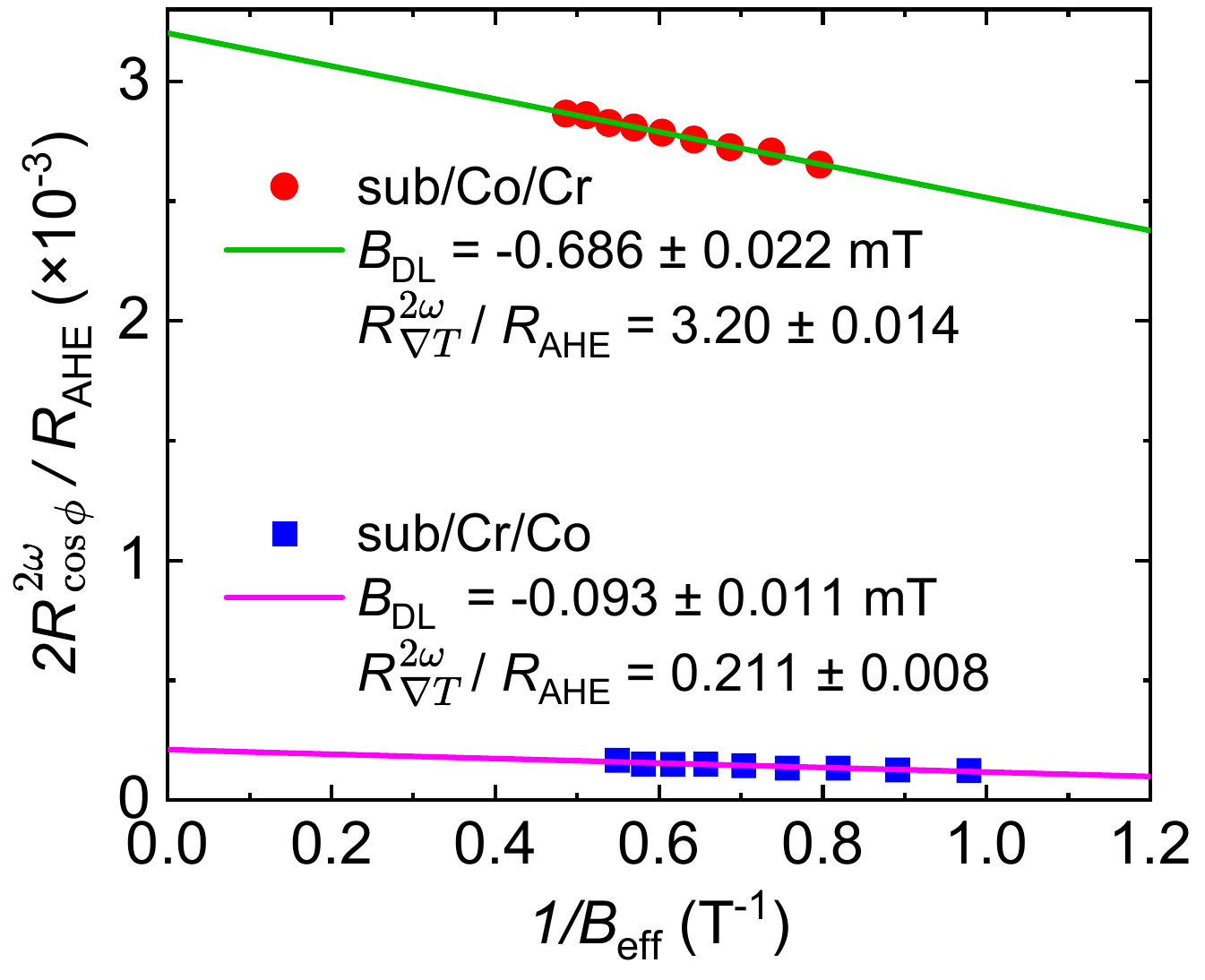}\label{fig:R2w_vs_1overBeff}}\hspace{0.1cm}
\subfloat{\xincludegraphics[height=0.25\textwidth,label=(f)]{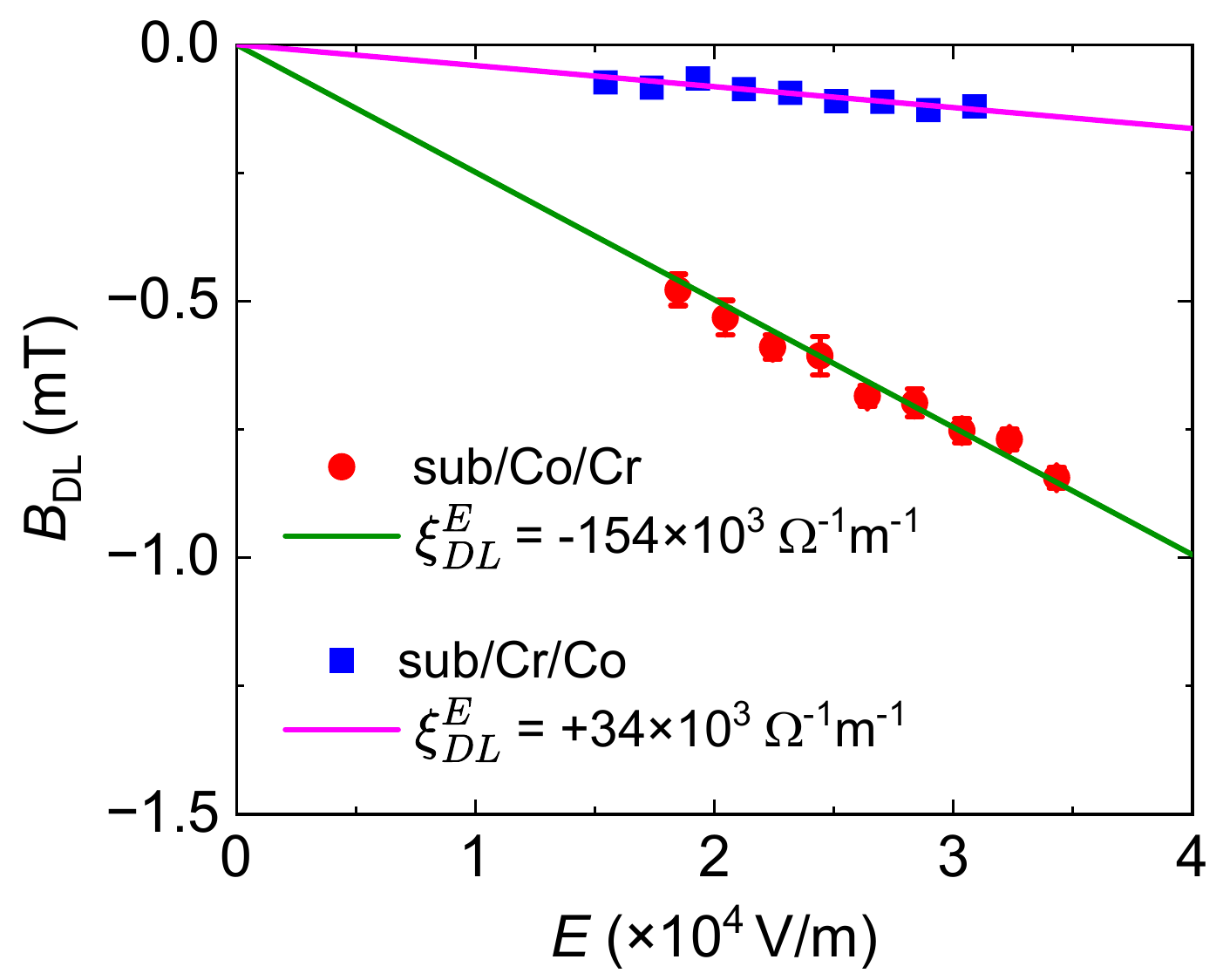}\label{fig:B_DL_vs_E}}\hspace{0.15cm}\vspace{-0.3cm}
\caption{
  (a) Schematic of the harmonic Hall measurement.
  The angular dependence of the second harmonic Hall resistance for (b) sub/Co(2)/Cr(10) and (c) sub/Cr(10)/Co(2) samples.
  The solid black lines are fits with eq.(\ref{eq:avci2014}).
  (d) Schematic of the effective damping-like field $B_{\text{DL}}$ in a FM/NM bilayer and a NM/FM bilayer with the reversed stacking order.
  (e) $2 R^{2\omega}_{\cos{\phi}}/R_{\text{AHE}}$ as a function of $1/B_{\text{eff}}$ and (f) $B_{\text{DL}}$ as a function of the applied electric field for sub/Co(2)/Cr(10) and sub/Cr(10)/Co(2).
  }
  \vspace{-0.2cm}
\end{figure*}

\section{Experiment}

The Cr/Co films are deposited at room temperature on insulating Si(100) substrates with 300\,nm of thermal SiO$_2$.
The deposition is performed by magnetron sputtering in a chamber with a base pressure of $5 \times 10^{-8}$\,Torr. 
To protect the films from oxidation, they are capped with SiO$_2$\,(10\,nm), except for sub/Cr/Co/Cr/SiO$_2$/Cr samples in which case 1--2\,nm of self-limiting native Cr oxide serves as a protective layer~\cite{allen_x-ray_1978,palacio_aes_1987,maurice_xps_2000}.
The films are patterned into 5\,$\mu$m wide Hall cross devices by photolithography and argon ion milling.

Using the harmonic Hall technique, we measure the damping-like SOT.
A sinusoidal current at a frequency of 987\,Hz is applied to the fabricated devices.
The voltage across the Hall probes is passed to a lock-in amplifier to measure the first and second harmonic voltages. 
To acquire the angular dependence, the sample is rotated in the film plane while the external magnetic field is held constant.
The schematic of the experimental geometry is shown in Fig.~\ref{fig:HH_schematics}.
All measurements are performed at 300\,K.

\section{Results and Discussion}

In Fig.~\ref{fig:R2w_angular_scan_sub_Co_Cr} and~\ref{fig:R2w_angular_scan_sub_Cr_Co}, we show the second harmonic Hall resistance $R^{2\omega}_{xy}$ measured for sub/Co(2)/Cr(10) and sub/Cr(10)/Co(2) with the reversed layer order during the rotation of external magnetic field $B_{\text{ext}}$ in the $xy$-plane.
The layer thickness in parentheses is given in nm.
The angle between the directions of the current and magnetic field is denoted as $\phi$.
All $R^{2\omega}_{xy}$ data is symmetrized relative to $\phi = 180\degree$ to exclude artificial signals due to a misalignment between the external magnetic field and the film plane~\cite{avci_interplay_2014}.
In the standard HH analysis~\cite{avci_interplay_2014}, measuring the angular dependence of $R^{2\omega}_{xy} (\phi)$ for different strengths of the magnetic field, strong enough to saturate the magnetization along $B_{\text{ext}}$, allows to separate the damping-like SOT (DL-SOT) contribution and the undesired thermal effects.
$R^{2\omega}_{xy} (\phi)$ is expressed as~\cite{avci_interplay_2014}:
\begin{equation}
\begin{split}
    R^{2\omega}_{xy} (\phi) & = \dfrac{1}{2} \left( R_{\text{AHE}} \dfrac{B_{\text{DL}}}{B_{\text{ext}}+B_{\text{k}}} + R^{2\omega}_{\nabla T} \right)\, \cos{\phi} \\  
    & + R_{\text{PHE}} \dfrac{B_{\text{FL}}+B_{\text{Oe}}}{B_{\text{ext}}} (2\cos^{3}{\phi}-\cos{\phi}) \\ & = R^{2\omega}_{\cos{\phi}} \cos{\phi} + R^{2\omega}_{\cos{3\phi}} (2\cos^3{\phi}-\cos{\phi}) \,,
    \label{eq:avci2014}
\end{split}
\end{equation}
where $R_{\text{AHE}}$ and $R_{\text{PHE}}$ are the first harmonic anomalous Hall and planar Hall resistances,
$B_{\text{DL}}$ and $B_{\text{FL}}$ are the damping-like and field-like effective fields, $B_{\text{k}}$ is the effective demagnetizing magnetic field of the ferromagnetic layer; $B_{\text{Oe}}$ is the current-induced Oersted field; 
$R^{2\omega}_{\nabla T}$ is the second harmonic thermal resistance, a sum of contributions from the anomalous Nernst effect (ANE) and
spin Seebeck (SSE)
generated by an out-of-plane temperature gradient, $\nabla T$, due to the Joule heating.
We note that in early papers a factor of 1/2 is missing on the right-hand side of the eq.\,(\ref{eq:avci2014}), which might have caused some discrepancies in the reported values of spin-orbit torque~\cite{lau_comments_2026}.
The contribution from the ONE is assumed to be negligibly small and is not included in eq.\,(\ref{eq:avci2014}).

The first term in eq.\,(\ref{eq:avci2014}) is proportional to $\cos{\phi}$ and holds the information about the DL-SOT, the second term is proportional to $(2\cos^3{\phi}-\cos{\phi})$ and holds the information about the FL-SOT. 
Fitting the measured angular dependence $R^{2\omega}_{xy} (\phi)$ with $R^{2\omega}_{\cos{\phi}} \cos{\phi} + R^{2\omega}_{\cos{3\phi}} (2\cos^3{\phi}-\cos{\phi}) $ and plotting the extracted parameter $R^{2\omega}_{\cos{\phi}}$ as a function of $1/B_{\text{eff}} = 1/(B_{\text{ext}}+B_{\text{k}})$ allows to separate DL-SOT and the thermoelectric effects.
The slope of the linear fit in Fig.~\ref{fig:R2w_vs_1overBeff} corresponds to $B_{\text{DL}}$, while the intercept is proportional to the thermal contribution normalized by anomalous Hall resistance, $R^{2\omega}_{\nabla T}/R_{\text{AHE}}$.
We note that in our samples, e.g. sub/Co(2)/Cr(10), the planar Hall resistance $R_{\text{PHE}}$ $\sim$1\,m$\Omega$ is too small to analyze $B_{\text{FL}}$, see Supplemental Material (SM) Section~S1~\cite{SupplMat}.
In the following, we focus on the analysis of $B_{\text{DL}}$.

Comparing the results for sub/Co/Cr vs sub/Cr/Co shown in Fig.~\ref{fig:R2w_vs_1overBeff}, we observe that in sub/Co/Cr $B_{\text{DL}}$ is several times larger than in sub/Cr/Co.
Furthermore, the sign of $B_{\text{DL}}$ stays the same upon inversion of the stacking sequence.
This is unexpected since $\bm B_{\text{DL}} \sim [\bm{m} \times \bm{\sigma}]$, where $\bm m$ is magnetization of FM layer and $\bm \sigma$ is the spin polarization.
If $\bm \sigma$ is generated by the spin Hall effect (SHE) in NM layer, then it has the opposite sign on the top and bottom surface of the NM, therefore $B_{\text{DL}}$ should have the opposite sign for NM/FM vs FM/NM stacking, as illustrated in Fig.~\ref{fig:NM_FM_vs_FM_NM_schematics}.
The same sign of $B_{\text{DL}}$ cannot be explained by spin Hall effect, Rashba-Edelstein effect, or their orbital counterparts~\cite{kontani_giant_2009,go_intrinsic_2018,jo_gigantic_2018,salemi_orbitally_2019,ding_harnessing_2020,ding_observation_2022,choi_observation_2023,lyalin_magneto-optical_2023}, since all of them have the symmetry discussed above.
We also observe that together with $B_{\text{DL}}$ the thermal contribution $R^{2 \omega}_{\nabla T}/R_{\text{AHE}}$ 
is an order of magnitude larger in sub/Co/Cr than in sub/Cr/Co.
It raises the question whether the values of DL-SOT measured by harmonic Hall technique in Cr/Co bilayers are dominated by thermal artifacts and whether the assumption of the $R^{2 \omega}_{\nabla T}$ being independent of magnetic field holds true.
Additional measurements on sub/Co/Pt vs sub/Pt/Co are shown in SM Sec.~S2, demonstrating that in Pt/Co bilayers the sign of $B_{\text{DL}}$ reverses upon the reversal of the stacking order, as expected.

To account for different resistivity, saturation magnetization, and stacking order we convert $B_{\text{DL}}$ into a spin-orbit torque efficiency per applied electric field~\cite{nguyen_spin_2016}:
\begin{equation}
     \xi_{DL}^E = \dfrac{2e}{\hbar} M_s t_{FM} \dfrac{B_{DL}}{E}\eta\,,
    \label{eq:xi_E}
\end{equation}
where $e$ is the electron charge, $\hbar$ is the reduced Planck constant, $t_{FM}$ is the thickness of the Co layer, $M_s$ is its saturation magnetization, $E$ is the applied electric field,
$\eta$ represents the stacking sequence:
$\eta = 1$ for sub/FM/NM and $\eta = -1$ for sub/NM/FM.
$M_s$ of the samples is measured by superconducting quantum interference device magnetometry, $M_s = 1020 \pm 110$\,kA/m for sub/Co(2)/Cr(10) and $M_s = 1370 \pm 150$\,kA/m for sub/Cr(10)/Co(2).
The resistivities of sub/Co(2)/Cr(10) and sub/Cr(10)/Co(2) bilayers are 790\,$\Omega\,\text{nm}$ and 460\,$\Omega\,\text{nm}$, respectively.
The current dependence recalculated to the applied electric field is shown in Fig.~\ref{fig:B_DL_vs_E}. 
Using parameters from the linear fitting, we obtain $\xi_{DL}^E = (-154 \pm 20) \times 10^3\,\Omega^{-1} \text{m}^{-1}$ for sub/Co/Cr and $\xi_{DL}^E = (+34 \pm 5) \times 10^3\,\Omega^{-1} \text{m}^{-1}$ for sub/Cr/Co.
The opposite sign between $B_{DL}$ and $\xi_{DL}^E$ in sub/Cr/Co arises from the stacking-order convention in eq.\,(\ref{eq:xi_E})
The large negative $\xi_E$ value for sub/Co/Cr agrees with those reported in Ref.~\cite{sala_giant_2022} where the same measurement technique, stacking order, and thicknesses of the Co/Cr layers were used, while the positive $\xi_E$ value for sub/Cr/Co disagrees with the sign of the SHE in Cr~\cite{du_systematic_2014,qu_inverse_2015,lee_efficient_2021}.

\begin{figure}[t!]
\makebox[0.03cm]{}
\subfloat{\xincludegraphics[height=0.25\textwidth,label=(a)]{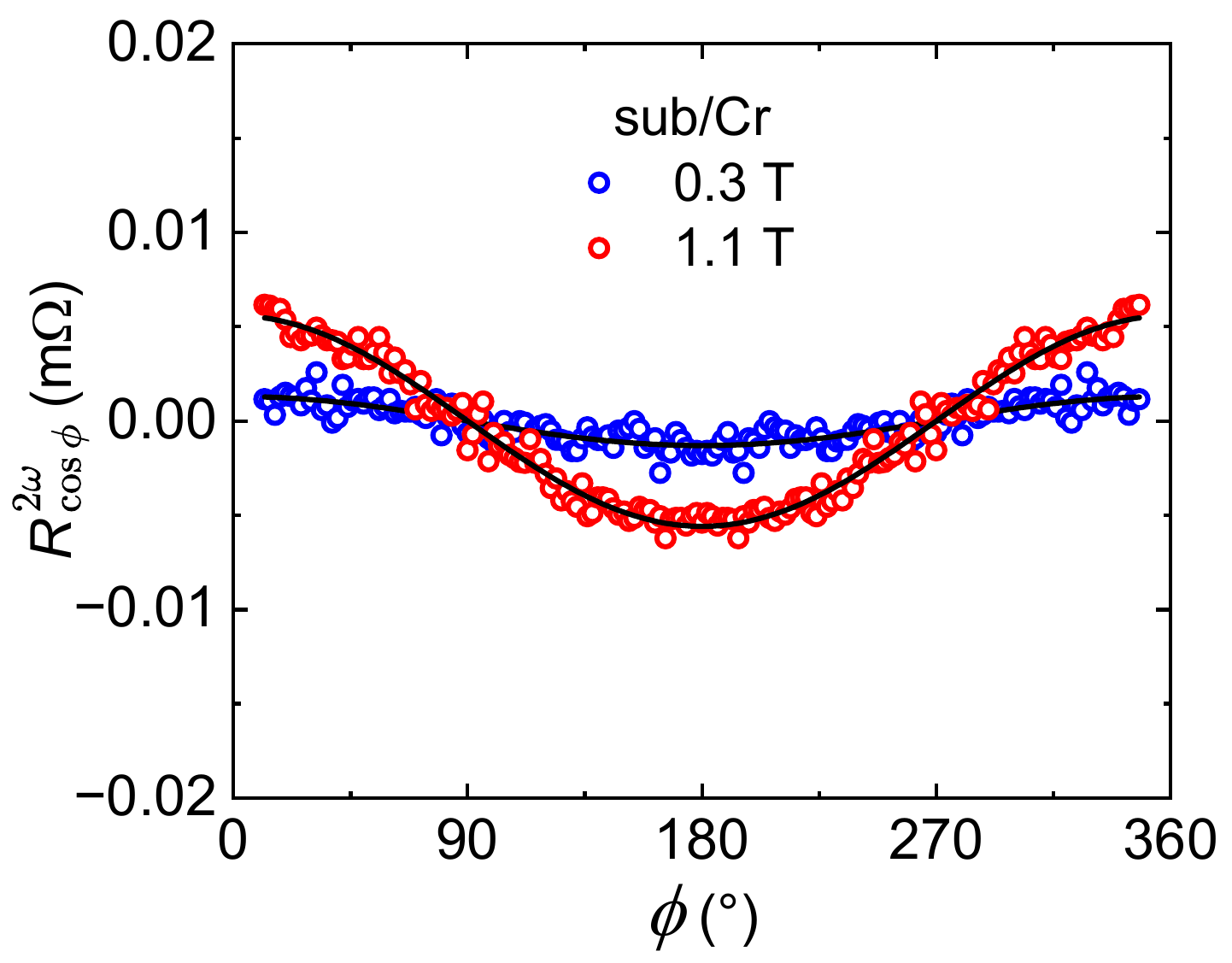}\label{fig:R2w_Cr}}\vspace{-0.4cm}
\makebox[0.0cm]{}
\subfloat{\xincludegraphics[height=0.25\textwidth,label=(b)]{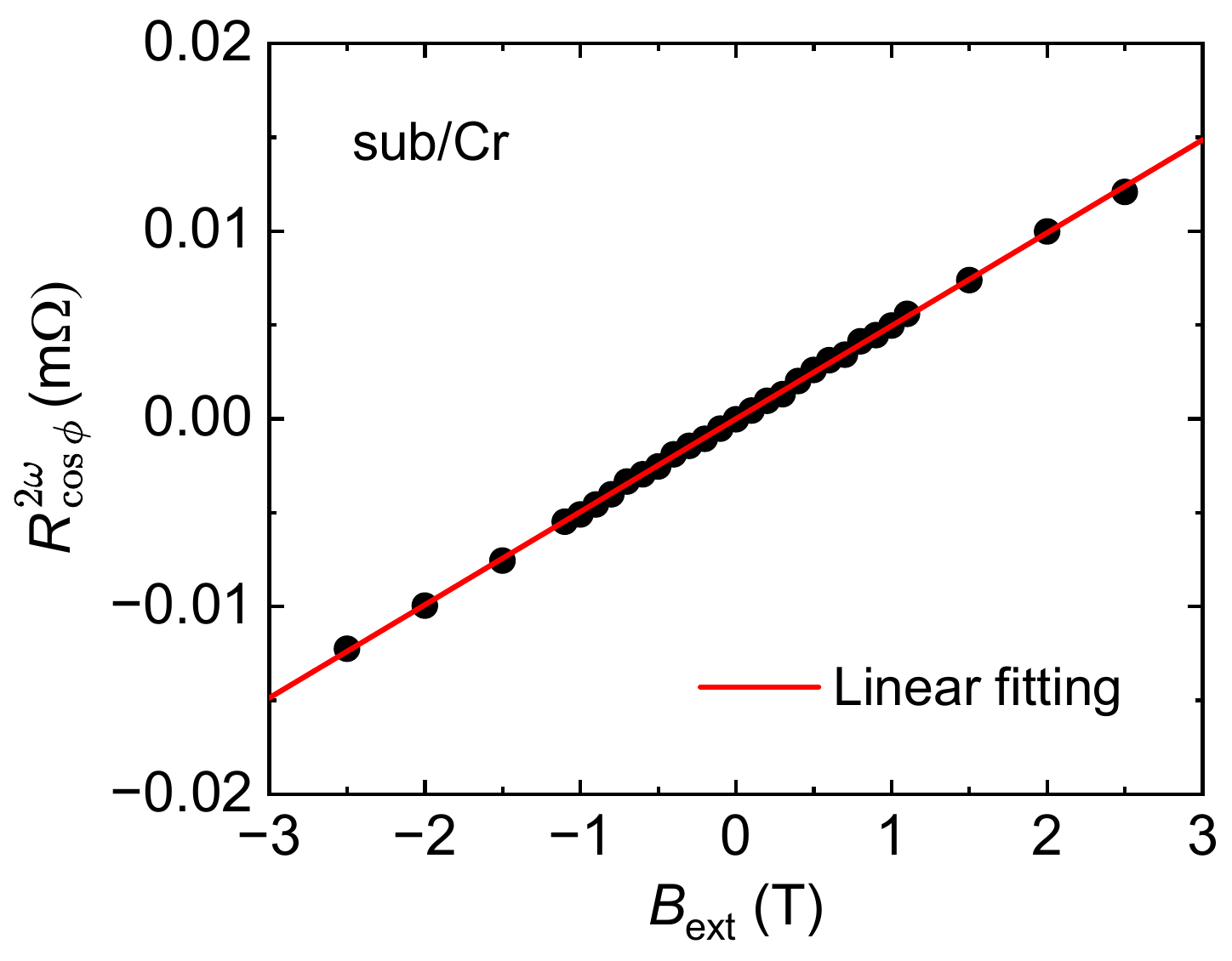}\label{fig:R2w_vs_Bext_Cr}}
\vspace{-0.4cm}
\makebox[0.04cm]{}
\subfloat{\xincludegraphics[height=0.25\textwidth,label=(c)]{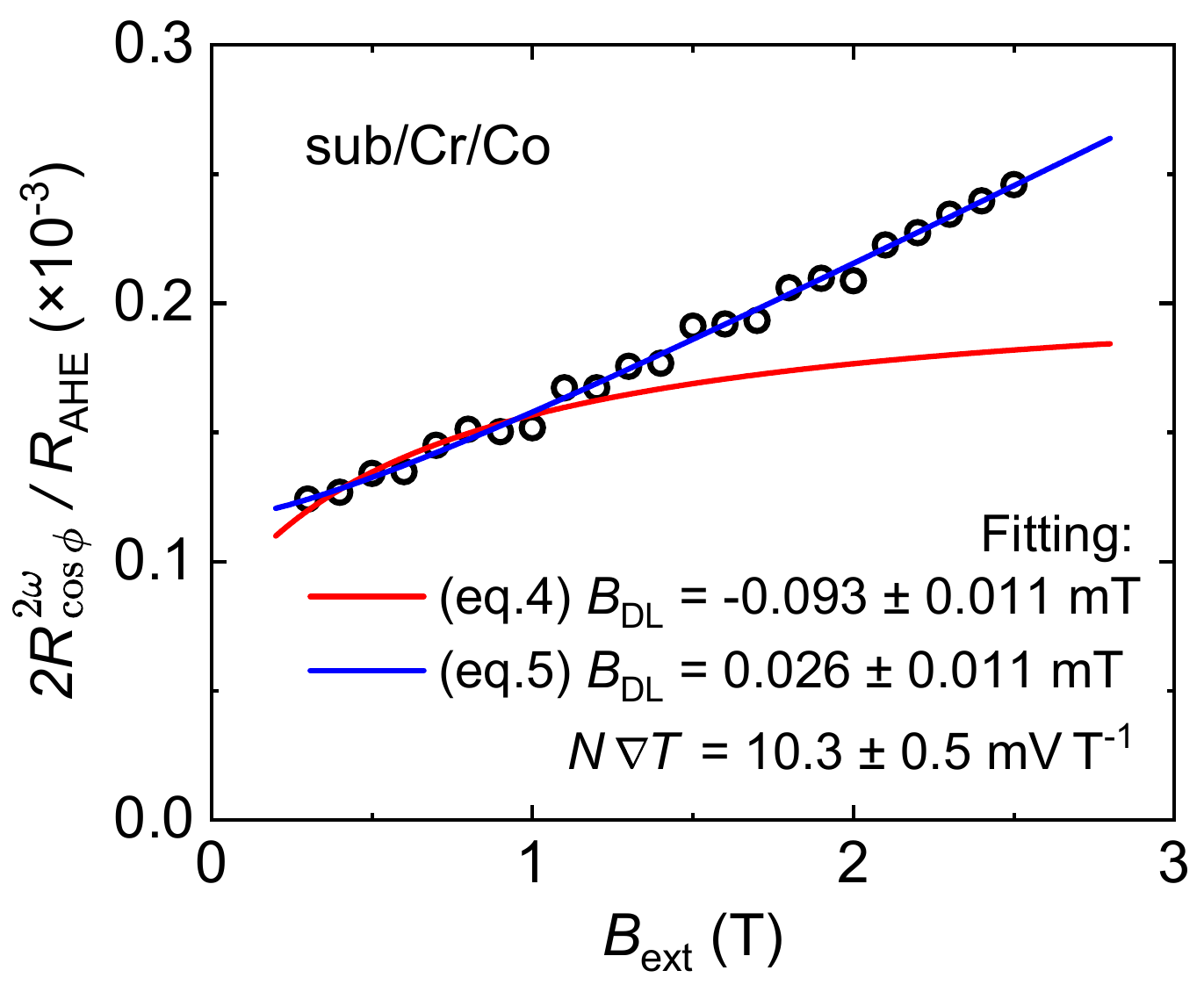}\label{fig:different_fittings_sub_Cr_Co}}\vspace{-0.3cm}
\caption{
  (a) Angular dependence of the second harmonic Hall resistance for sub/Cr(12).
  (b) $R^{2\omega}_{\cos{\phi}}$ as a function of $B_{\text{ext}}$ for sub/Cr(12).
  (c) Fitting of $R^{2\omega}_{\cos{\phi}}(B_{\text{ext}})$ data for sub/Cr(10)/Co(2) with eq.\,(\ref{eq:fitting_without_ONE}) in 0.3 -- 1.1\,T range (without the ONE) and with eq.\,(\ref{eq:fitting_with_ONE}) in 0.3 -- 2.5\,T range (with the ONE).
  }
  \vspace{-0.2cm}
\end{figure}

To find the origin of such large discrepancy between sub/Co/Cr and sub/Cr/Co, we perform measurements on a single Cr layer {\it{without}} a FM layer, thus {\it{no}} SOT.
Fig.~\ref{fig:R2w_Cr} shows the angular dependence of the second harmonic resistance for sub/Cr(12) at $B_{\text{ext}} = 0.3$\,T
and 1.1\,T.
Surprisingly, we still observe a signal with a symmetry of $\cos{\phi}$, as if the DL-SOT is present.
For FM/NM bilayers, the difference between the angular scans of $R^{2\omega}_{xy}$ at different magnetic fields is a measure of the DL-SOT strength. 
One notes the $R^{2\omega}_{xy}$ difference in sub/Cr(12) for 0.3\,T
and 1.1\,T scans is comparable or even larger than the difference for sub/Cr(10)/Co(2) sample shown in Fig.~\ref{fig:R2w_angular_scan_sub_Cr_Co}, plotted with the same scale.
However, this difference cannot be assigned to the SOT for there is {\it{no}} FM layer.
Measuring the magnetic field dependence, Fig.~\ref{fig:R2w_vs_Bext_Cr}, we find that $R^{2\omega}_{\cos{\phi}}$ scales linearly with $B_{\text{ext}}$.
These angular and magnetic field dependencies provide telltale indications of the ordinary Nernst effect~\cite{roschewsky_spin-orbit_2019,chi_spin_2020}, 
which is considered to be negligible and consequently omitted in the HH analysis in metallic FM/NM bilayers~\cite{avci_interplay_2014}.

To account for the ONE, an additional term needs to be added to the right-hand side of eq.\,(\ref{eq:avci2014}):
\begin{equation}
\dfrac{1}{2} R^{2\omega}_{\text{ONE}} = \dfrac{1}{2} \dfrac{w}{I_0} N \nabla T B_{\text{ext}} \cos{\phi} \,,
\label{eq:R2w_ONE}
\end{equation}
where $w$ is the Hall cross width, $N$ is the ordinary Nernst coefficient of the FM/NM bilayer, $\nabla T$ is a temperature gradient along the film normal in the bilayer.
In a simple short-circuit model~\cite{xu_scaling_2008}:
\begin{equation*}
\begin{split}
N \nabla T & = \dfrac{R_{\text{FM}}}{R_{\text{NM}}+R_{\text{FM}}} N_{\text{NM}} \nabla T_{\text{NM}} \\ 
& + \dfrac{R_{\text{NM}}}{R_{\text{NM}}+R_{\text{FM}}} N_{\text{FM}} \nabla T_{\text{FM}}\,,
\label{eq:N_nabla_T_short_circuit}
\end{split}
\end{equation*}
where $N_{\text{NM}}$ ($N_{\text{FM}}$), $\nabla T_{\text{NM}}$ ($\nabla T_{\text{FM}}$), $R_{\text{NM}}$ ($R_{\text{FM}}$) are the ordinary Nernst coefficient, temperature gradient, and sheet resistance of the NM (FM) layer.

The part of eq.\,(\ref{eq:avci2014}) proportional to $\cos{\phi}$ can be re-written as:
\begin{equation}
R^{2\omega}_{\cos{\phi}} = c_{\text{DL}}\dfrac{1}{B_{\text{ext}}+B_{\text{k}}} + c_{\text{ANE}} \,.
\label{eq:fitting_without_ONE}
\end{equation}
Adding the ONE term, eq.\,(\ref{eq:R2w_ONE}), to the HH analysis means that instead of eq.\,(\ref{eq:fitting_without_ONE}) 
the magnetic field dependence $R^{2\omega}_{\cos{\phi}} (B_{\text{ext}})$ should be fitted with an equation:
\begin{equation}
R^{2\omega}_{\cos{\phi}} = c_{\text{DL}} \dfrac{1}{B_{\text{ext}}+B_{\text{k}}} + c_{\text{ONE}} B_{\text{ext}} + c_{\text{ANE}} \,,
\label{eq:fitting_with_ONE}
\end{equation}
where, to shorten the notations, we denote fitting parameters that describe DL-SOT, ANE, and ONE contributions as $c_{\text{DL}}$, $c_{\text{ANE}}$, and $c_{\text{ONE}}$, respectively.
The fitting with eq.\,(\ref{eq:fitting_with_ONE}) vs  eq.\,(\ref{eq:fitting_without_ONE}), shown in Fig.~\ref{fig:different_fittings_sub_Cr_Co}, allows to correct the sign of the $B_{\text{DL}}$ to be negative, as expected for Cr/Co if the DL-SOT is dominated by the negative SHE in Cr.
It also allows to evaluate the average $N \times \nabla T$ of a bilayer system.
We will discuss values of $N$ and $\nabla T$ later.
We note that the fitting with eq.\,(\ref{eq:fitting_with_ONE}) for magnetic fields below 1\,T, a field range typical for this configuration of HH measurements~\cite{avci_interplay_2014}, cannot reliably separate the DL-SOT and ONE contributions, because the difference between $1/(B_{\text{ext}} + B_{\text{k}})$ and $B_{\text{ext}}$ functions is weak.
Therefore, measurements up to a few Teslas are required.
The corrected DL-SOT efficiency for sub/Cr/Co is $\xi_{DL}^E = (-10 \pm 4) \times 10^3 \,\Omega^{-1} \text{m}^{-1}$.
The measurements for sub/Co(2)/Cr(10) are shown in SM Sec.~S3
The corrected DL-SOT for sub/Co(2)/Cr(10) is estimated to be $\xi_{DL}^E = (-76 \pm 12) \times 10^3 \,\Omega^{-1} \text{m}^{-1}$.
We conclude that the standard HH analysis without the ONE term overestimates the DL-SOT in Cr/Co by at least 2 times.

The origin of the different DL-SOT in sub/Co/Cr vs sub/Cr/Co can be due to an actual SOT, for example a larger interfacial SOT at one of the interfaces,
or it can be due to the unaccounted thermal artifacts beyond the ONE.
In the first case, the different SOT can be explained by different crystal structure and interface quality of the samples, since both, Co and Cr, though grown with the same deposition parameters, are deposited on the different underlayers.
In the second case, the DL-SOT value measured on sub/Cr/Co, $\xi_{DL}^E = (-10 \pm 4) \times 10^3 \,\Omega^{-1} \text{m}^{-1}$, should be trusted more because for this stacking sequence the $R^{2\omega}_{xy}$ thermal background is $\sim$\,10 times smaller than for sub/Co/Cr.

\begin{figure}[b!]
\subfloat{\xincludegraphics[height=0.22\textwidth,label=(a)]{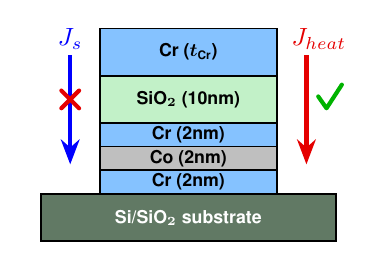}\label{fig:schematics_of_cr_co_cr_sio2_cr_stack}}\vspace{-0.6cm}\hfill
\subfloat{\xincludegraphics[height=0.4\textwidth,label=(b)]{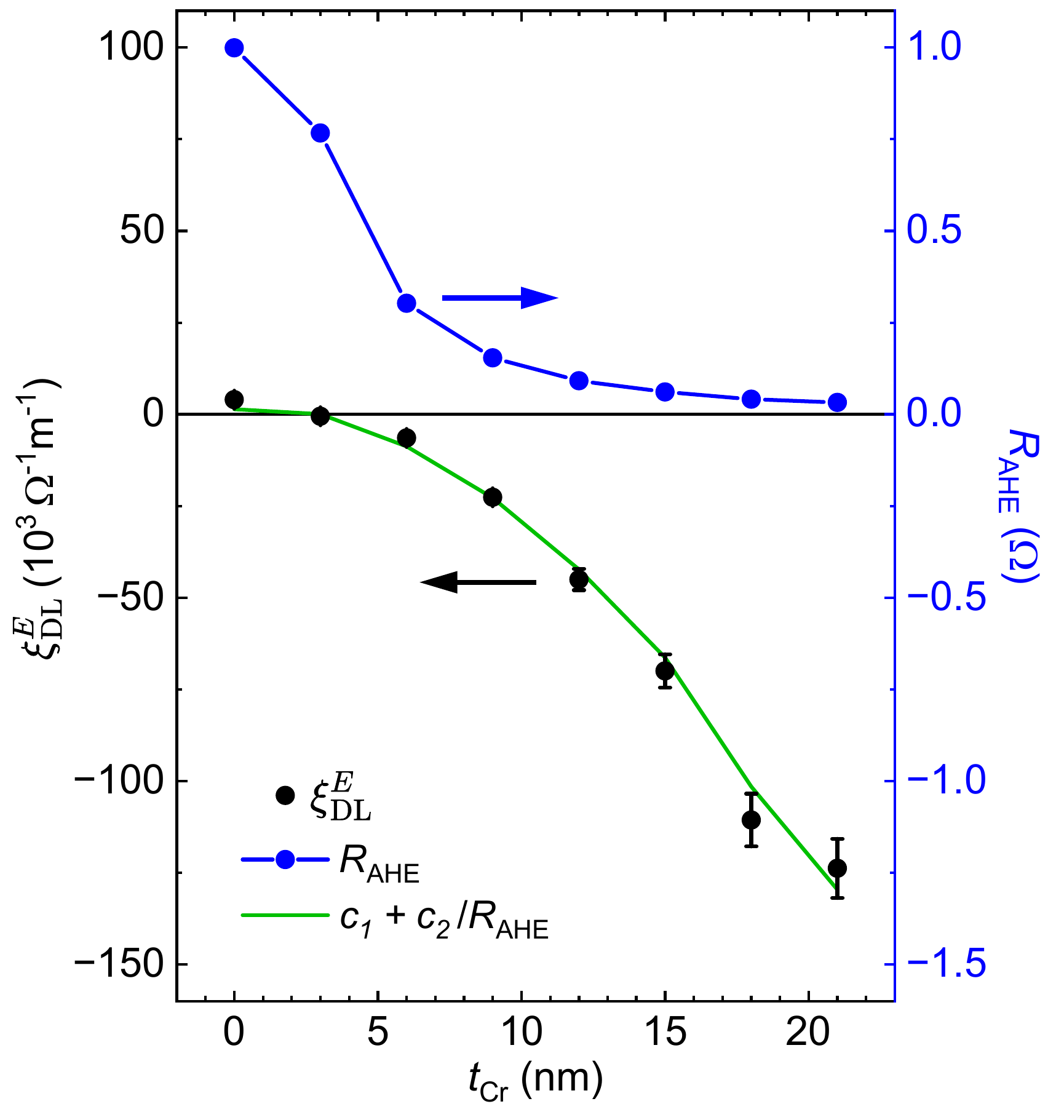}\label{fig:xi_E_for_the_multilayer_stack}}
\vspace{-0.2cm}\hfill
\caption{
  (a) Schematics of a control sample sub/Cr(2)/Co(2)/Cr(2)/SiO$_2$(10)/Cr($t_{\text{Cr}}$) designed to have {\it{zero}} SOT.
  (b) DL-SOT efficiency $\xi_{DL}^E$ and anomalous Hall resistance $R_{\text{AHE}}$ as a function of the top Cr layer thickness, $t_{\text{Cr}}$.
  }
  \vspace{-0.3cm}
\end{figure}

\begin{figure*}[ht!]
\subfloat{\xincludegraphics[height=0.25\textwidth,label=(a)]{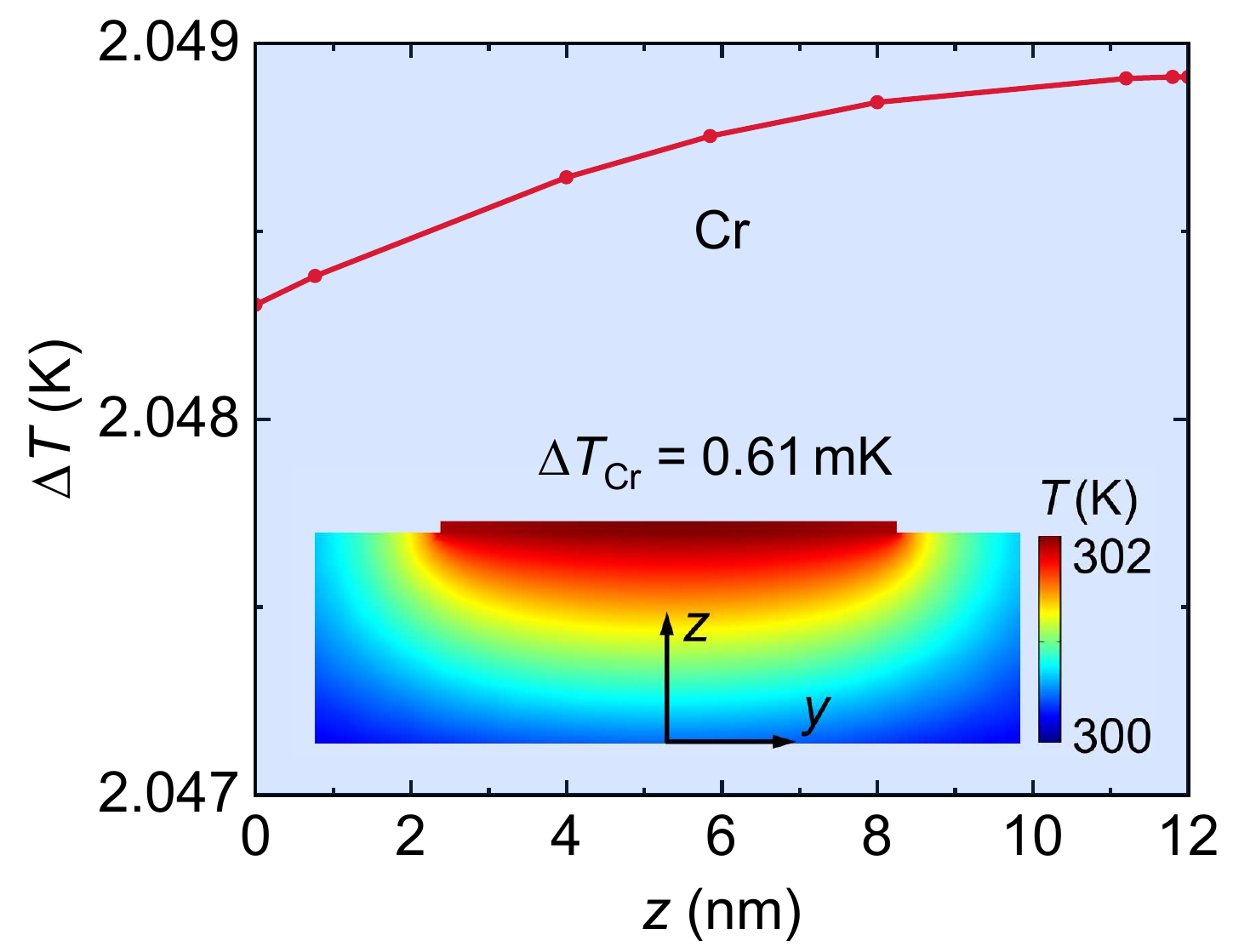}\label{fig:comsol_sub_cr}}\hfill
\subfloat{\xincludegraphics[height=0.25\textwidth,label=(b)]{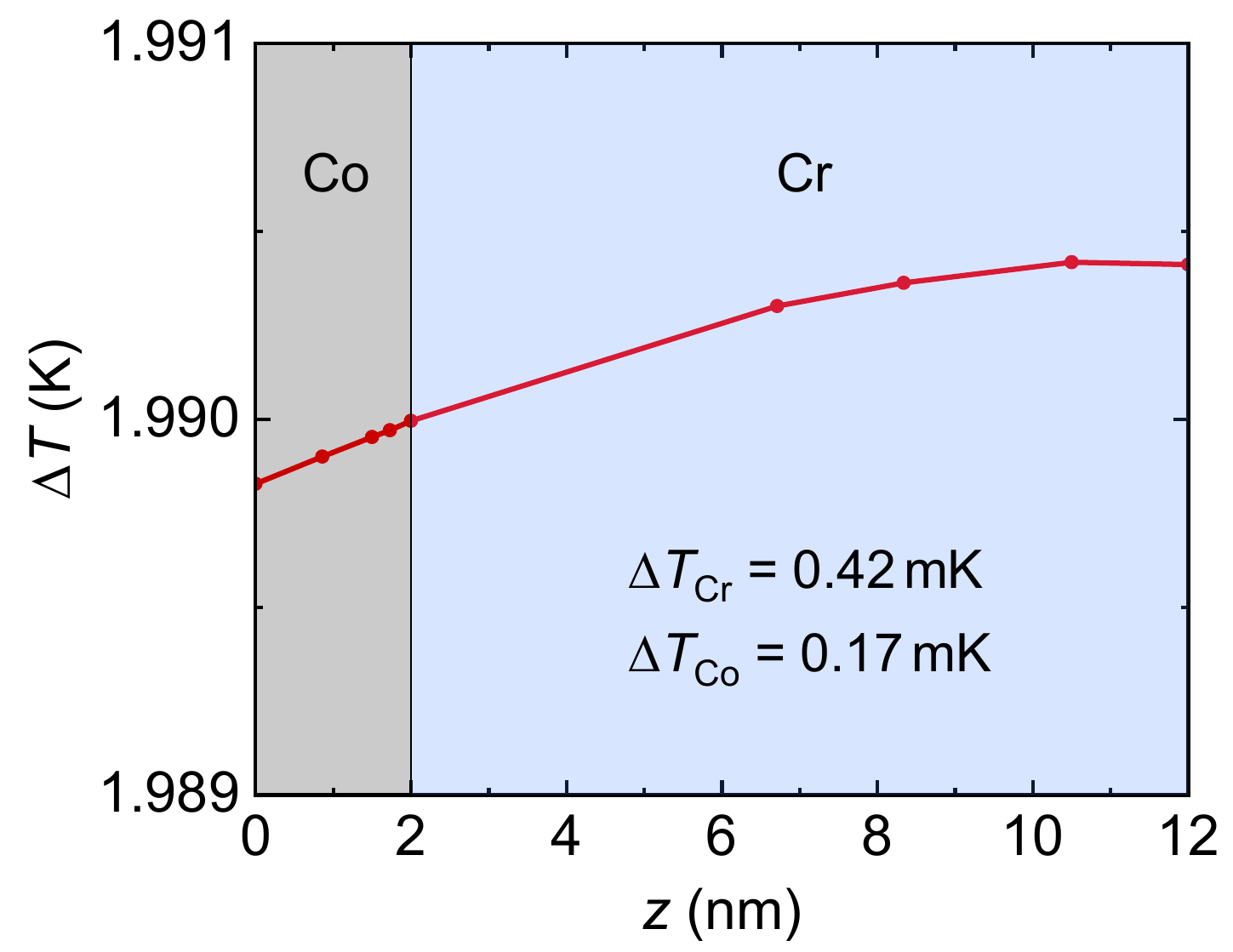}\label{fig:comsol_sub_co_cr}}\hfill
\subfloat{\xincludegraphics[height=0.25\textwidth,label=(c)]{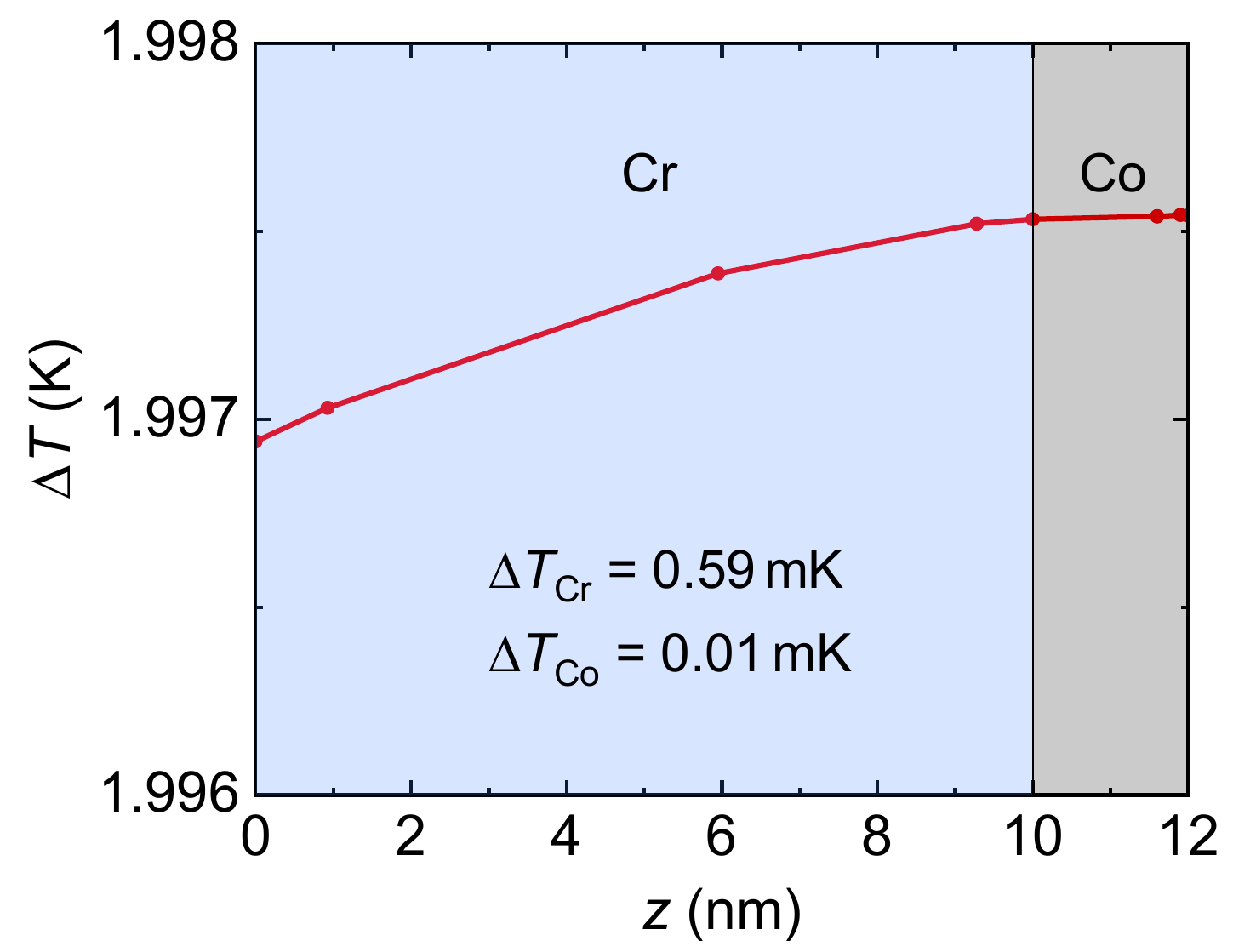}\label{fig:comsol_sub_cr_co}}
\vspace{-0.3cm}\hfill
\caption{
  COMSOL simulations of the out-of-plane temperature profile in (a) sub/Cr(12),
  (b) sub/Co(2)/Cr(10), (c) sub/Cr(10)/Co(2) for an applied electric field $E = 24$\,kV/m.
  Temperature is given relative to 300\,K at the backside of the substrate $\Delta T(z) = T(z) - 300$\,K. 
  The interface with the substrate is $z = 0$\,nm, the interface with the air is $z = 12$\,nm.
  The inset in (a) shows 2D temperature distribution in $yz$-plane of the device.  
  }
\label{fig:COMSOL_all_figures}
  \vspace{-0.3cm}
\end{figure*}

To better understand the magnitude of the spurious $\xi^E_{\text{DL}}$ values generated by the ONE, we perform HH measurements on a series of sub/Cr(2)/Co(2)/Cr(2)/SiO$_2$(10)/Cr($t_{\text{Cr}}$) 
multilayers, which are designed to have {\it{zero}} SOT.
The Cr/Co/Cr part of the stack is symmetrical by design to cancel out any possible spin-orbit torque contributions.
As illustrated in Fig.~\ref{fig:schematics_of_cr_co_cr_sio2_cr_stack}, the insulating SiO$_2$(10) layer blocks spin and orbital currents generated by the top Cr layer, while the out-of-plane heat flow is still present and thermoelectric effects can contribute to the second harmonic voltage.
In the data analysis we still use eq.\,(\ref{eq:avci2014}) that does not account for the ONE.
The first data point in Fig.~\ref{fig:xi_E_for_the_multilayer_stack}, $t_{\text{Cr}}=0$\,nm, confirms that without top Cr layer $\xi_{DL}^E$ in Cr/Co/Cr is close to zero.
However, as $t_{\text{Cr}}$ increases, the $\xi_{DL}^E$ monotonically decreases reaching values of  $-10^5$\,$\Omega^{-1}$\,m$^{-1}$ for $t_{\text{Cr}} \gtrsim 18$\,nm.
The $\xi_{DL}^E (t_{\text{Cr}})$ behavior is well described by equation $c_1 + c_2/R_{\text{AHE}}$, shown by the solid green line in Fig.~\ref{fig:xi_E_for_the_multilayer_stack}, where a constant $c_1 \approx 0$ represents a true value of $\xi_{DL}^E$ in Cr/Co/Cr, while $c_2$ represents a small unaccounted effect, e.g., the ONE.
As follows from eq.\,(\ref{eq:avci2014})-(\ref{eq:R2w_ONE}), if the ONE is not properly considered than its contribution to $\xi_{DL}^E$ is magnified by a factor $1/R_{\text{AHE}}$.
Due to the current shunting and short-circuit effects~\cite{xu_scaling_2008},
$R_{\text{AHE}}$ decreases dramatically with increasing thickness of the top Cr layer, from 1\,$\Omega$ at $t_{\text{Cr}}=0$ to 33\,m$\Omega$ at $t_{\text{Cr}}=21$\,nm, see Fig.~\ref{fig:xi_E_for_the_multilayer_stack}.
As a result, the ONE term can play a dominant role in harmonic Hall measurements generating significant spurious values of $\xi_{DL}^E$ on the order of $10^4 - 10^5$\,$\Omega^{-1}$\,m$^{-1}$.

Next, we model temperature gradient in Cr/Co bilayers using COMSOL simulations, discuss how stacking sequence affects the thermal gradient, and estimate the Nernst coefficient in Cr.
We simplify the Hall cross geometry of the device to a $2\times 0.5\,\mu$m lateral strip of Cr/Co on a $10\times10\times0.3\,\mu$m slab of SiO$_2$, which is on the top of a $10\times10\times2\,\mu$m slab of Si.
We keep the thicknesses of Cr, Co, and SiO$_2$ layers as in the experiment, while decreasing the lateral dimensions by 10 times to reduce the number of finite-size elements in the simulations.
We assume no heat conduction at the top surface of the sample, due to the low thermal conductivity of the air, $\kappa_{\text{air}} = 0.026$\,W m$^{-1}$ K$^{-1}$, while fixing the temperature at the backside of the substrate to $T = 300$\,K.
We use the following bulk values of the thermal conductivities:
$\kappa_{\text{Si}} = 130$\,W m$^{-1}$ K$^{-1}$, $\kappa_{\text{SiO$_{2}$}} = 1.3$\,W m$^{-1}$ K$^{-1}$, $\kappa_{\text{Cr}} = 94$\,W m$^{-1}$ K$^{-1}$, $\kappa_{\text{Co}} = 100$\,W m$^{-1}$ K$^{-1}$.
The Joule heating generated in Cr/Co bilayers is calculated using the device geometry, a typical applied electric field $E =
24$ kV/m, and experimental resistivity of  $\rho_{\text{Cr}}$ = 480\,$\Omega$\,nm.
The resistivity of Co is assumed to be equal, $\rho_{\text{Co}} = \rho_{\text{Cr}}$.

Using the bulk values of thermal conductivity for Cr and Co leads to an underestimation of the temperature gradients in these layers by a factor of $\kappa_{\text{bulk}}/\kappa_{\text{film}}$.
For the thin metal films this factor is typically larger than 1
because the additional electron scattering at the interfaces and grain boundaries reduces the thermal conductivity of the film. 
However, Wiedemann-Franz law, $\kappa = L T \sigma$, where L is Lorentz number and $\sigma$ is electrical conductivity, cannot be applied to estimate $\kappa_{\text{film}}$, because for thin films the phonon contribution to thermal conductivity starts playing a significant role~\cite{islam_kappa_W_Ru_2025}.
For films in the range of 0--40\,nm, some $\kappa_{\text{bulk}}/\kappa_{\text{film}}$ room temperature values reported in the literature are: $\sim$3 for Cu~\cite{lugo_kappa_Cu_2016,islam_kappa_Cu_2024}, $\sim$3 for Au~\cite{mason_kappa_Au_2020}, $\sim$3 for W~\cite{islam_kappa_W_Ru_2025}, $\sim$2 for Ru~\cite{islam_kappa_W_Ru_2025}, $\sim$2 for Ir~\cite{lin_kappa_Ir_2013}.
Furthermore, similar to electrical conductivity, the thermal conductivity is sensitive to the growth conditions of the films.
Thus, we do not attempt to explicitly correct for $\kappa_{\text{bulk}}/\kappa_{\text{film}}$ factor and use the bulk values of the thermal conductivity in the simulations, though understanding that this leads to $\kappa_{\text{bulk}}/\kappa_{\text{film}}$ times smaller $\nabla T$.

The results of 3D COMSOL simulations for sub/Cr, sub/Co/Cr, sub/Cr/Co are shown in Fig.~\ref{fig:COMSOL_all_figures}.
The solution of 1D heat equation, $\kappa\, \partial^2 T / \partial^2 z = \dot{q}$, for the temperature profile $T(z)$ is a quadratic function $T \sim$\,$z^2$ with a maximum at the top interface due to the boundary condition of no heat conduction $\partial T/\partial z = 0$, where
$\dot{q}$ is Joule heating power density.
The solution of the 1D heat equation agrees well with the 3D results of COMSOL simulations and explains why the experimental $R^{2\omega}_{\nabla T}$ is an order of magnitude larger for sub/Co/Cr in comparison with sub/Cr/Co.
The main origin of $R^{2\omega}_{\nabla T}$ is the anomalous Nernst effect in the Co layer, which is proportional to $\nabla T_{\text{Co}}$. 
Due to the $T(z)$ being a quadratic function, the temperature drop in Co, $\Delta T_{\text{Co}}$, is much larger when Co is the bottom vs top layer, $\Delta T_{\text{Co}} = 0.17$\,mK vs 0.01\,mK.
This explanation differs from the one given in Ref.~\cite{avci_interplay_2014} that focuses on the Joule heating generated by more conductive vs less conductive layer in the FM/NM stack, which does play a role in determining the temperature gradient $\nabla T_{\text{FM}}$,
however $T(z)$ profile is more important.
This leads to an important conclusion that in order to minimize thermal artifacts due to the ANE in HH measurements the stacking order with FM layer on the top is preferable.
Otherwise, a large thermal background due to the ANE can be present, as seen in Fig.~\ref{fig:R2w_angular_scan_sub_Co_Cr} for sub/Co/Cr.
The deviation of this background from a constant, its drift and noise can serve as an additional source of error in determining the SOT.
It is also worth noting that, according to the heat equation, the temperature gradient in each layer is inversely proportional to its thermal conductivity, $\nabla T \sim 1/\kappa$.

\begin{table*}[ht]
    \centering
    \begin{tabular}{ l  c  c  c  c  r }
        \toprule
        \textbf{Material} & \textbf{Hall Coeff.} & \textbf{Seebeck Coeff.} & \textbf{Measured} & \textbf{Estimated} & \textbf{Thermal}\\
        & $\phantom{-}$ $R_{\text{H}}$ (10$^{-11}$ m$^3$ C$^{-1}$) $\phantom{-}$ & $S$ ($\mu$V\,K$^{-1}$) & \textbf{Nernst Coeff.} & \textbf{Nernst Coeff.} & \textbf{Conductivity} \\
        & & & $N$ (nV\,K$^{-1}$\,T$^{-1}$) & $\phantom{-}N = 2/3\, S R_H \sigma\phantom{-}$ & $\kappa$ (W m$^{-1}$ K$^{-1}$)\\
        \midrule
        Pt & $-2$~\cite{hurd_hall_1972} & $-5$~\cite{moore_absolute_1973} & 13~\cite{luck_zur_1964}, $9$ (f)~\cite{chiang_nernst_2024}, 20 (f,max) [*] & $0.6$ & $72$\\
        Pd & $-8$~\cite{hurd_hall_1972} & $-10$~\cite{luck_zur_1964} & 40~\cite{luck_zur_1964} & $5$ & $72$\\
        Cu & $-5$~\cite{hurd_hall_1972} & $1.8$~\cite{roberts_absolute_1981} & $-20$ ($+900$ at 20\,K)~\cite{fletcher_nernst-ettinghausen_1972} & $-3.5$ & $401$\\
        Au & $-7$~\cite{hurd_hall_1972} & $2.0$~\cite{roberts_absolute_1981} & $-16$ ($+180$ at 20\,K)~\cite{fletcher_nernst-ettinghausen_1972} & $-4.3$ & $318$ \\
        Co & $-8$~\cite{hurd_hall_1972} & $-30$~\cite{laubitz_transport_1973} & $-$ & $27$ & $100$ \\
        Ni & $-6$~\cite{hurd_hall_1972} & $-19$~\cite{laubitz_transport_1976} & $-$ & $11$ & $91$ \\
        Gd & $-100\,(H||ab)$~\cite{lee_hall_1967} & $-2.5$~\cite{sill_seebeck_1965} & $-$ & $1.3$ & $11$ \\
        & $-520\,(H||c)$~\cite{lee_hall_1967} &  $-1$~\cite{sill_seebeck_1965} & $-$ & $2.5$ & \\
        Cr & $38$~\cite{hurd_hall_1972} & $22$~\cite{moore_thermal_1977} & $120$ (f,max) [*] & 44 & 94 \\
        \bottomrule
    \end{tabular}
    \caption{Reported ordinary Hall coefficient, Seebeck coefficient, ordinary Nernst coefficient, estimated Nernst coefficient, and thermal conductivity of selected bulk materials at room temperature, unless otherwise specified. (f) -  studied samples are thin films, (max) - the upper limit, [*] - this work.}
    \label{tab:material_properties}
    \vspace{-0.3cm}
\end{table*}

Using $\Delta T_{\text{Cr}} = 0.61$\,mK in a single layer of 12\,nm thick Cr, see Fig.~\ref{fig:comsol_sub_cr} for the calculated $T(z)$ profile, the slope of $R^{2\omega}_{xy}$ vs $B_{\text{ext}}$ from Fig.~\ref{fig:R2w_vs_Bext_Cr}, and equation:
$$R^{2\omega}_{\cos{\phi}} = \dfrac{1}{2} \dfrac{w}{I_0} N_{\text{Cr}} B_{\text{ext}} \dfrac{\Delta T_{\text{Cr}}}{t_{\text{Cr}}}\,\text{,}$$
we estimate the upper limit of the Nernst coefficient in Cr: $N_{\text{Cr}} < 120$\,nV\,K$^{-1}$\,T$^{-1}$.
A more realistic $N_{\text{Cr}}$ value might be 2--3 times smaller 
if the discussed $\kappa_{\text{bulk}}/\kappa_{\text{film}}$ factor is measured and taken into account, which is beyond the scope of this work.
To verify the validity of our approach, we conduct the HH measurements on sub/Pt (12\,nm) and perform COMSOL simulations for $\Delta T_{\text{Pt}}$, see SM Sec.~S4.
We estimate $N_{\text{Pt}} < 20$\,nV\,K$^{-1}$\,T$^{-1}$, which agrees well with the literature: 13\,nV\,K$^{-1}$\,T$^{-1}$ for bulk~\cite{luck_zur_1964}, 9\,nV\,K$^{-1}$\,T$^{-1}$ for 15\,nm thick films~\cite{chiang_nernst_2024}.

Unfortunately, the experimental reports on the ordinary Nernst coefficient in metals are scarce.
In Table~\ref{tab:material_properties} we provide a summary of Hall, Seebeck, and Nernst coefficients measured mostly on bulk materials.
We also provide an estimated value of the Nernst coefficient according to the formula that connects the Nernst, Hall, and Seebeck effects: $N = 2/3\, S R_{\text{H}} \sigma$, where $R_{\text{H}}$ is the Hall coefficient, $S$ is the Seebeck coefficient, $\sigma$ is the conductivity. 
However, this relation only holds true for a one-band system under an assumption of a linear dependence of the Hall angle on energy~\cite{behnia_nernst_2009}.
It correctly predicts the sign of $N$ for Pt, Pd, Cu, and Au, but underestimates its value.
As can be seen from Table~\ref{tab:material_properties}, our upper limit estimate for $N_{\text{Cr}}$ is
3--10 times larger than the Nernst coefficient measured in Pt, Pd, Cu, and Au at room temperature,
but smaller than Nernst coefficient in Cu and Au at low temperatures.
This comparison demonstrates that $N_{\text{Cr}}$ is not exceptionally large, but is comparable in magnitude to other metals.
Therefore, we argue that the ordinary Nernst effect should be carefully considered when using the harmonic Hall technique to quantify the SOT, not only in Cr-based systems, but in other metallic bilayers.
For example, recent studies on spin-orbit torque in CuO$_{\text{x}}$-based systems observed an enhancement of the SOT efficiency at low temperatures using the HH measurements~\cite{ding_orbital_2024,ding_generation_2025}.
Considering the large value of $N_{\text{Cu}}$ reported at low temperatures~\cite{fletcher_nernst-ettinghausen_1972}, 
the ONE could be contributing to the measured $\xi^E_{DL}$ values in those studies.

Based on the relatively large Hall and Seebeck coefficients in Co, there is a reason to expect $N_{\text{Co}}$ of the same sign and magnitude as in Cr, see Table~\ref{tab:material_properties}.
It is also important to emphasize, that it is not only Nernst coefficient that matters for harmonic Hall measurements, but the product $N \times \nabla T$.
As we have seen from the simulations, the temperature gradient depends on the stacking order.
Therefore, in addition to the ONE contribution from Cr layer, a contribution from the ONE in Co could be expected for sub/Co/Cr, while it should be $\sim$\,10 times smaller in sub/Cr/Co. 
Therefore, we argue that the HH measurements on a sub/NM/FM combined with the measurements on a reference sub/NM sample are more trustworthy than the measurements on a sub/FM/NM alone.
Additional measurements to assess the ONE contribution in Co are provided in SM Sec.~S5.

To illustrate the procedure with a reference sub/NM sample, we measure $R^{2\omega}_{xy}(\phi)$ data on a sub/Cr(12) in the 0.3--1.1\,T field range in which sub/Cr(10)/Co(2) is measured.
In this field range the difference between $1/(B_{\text{ext}} + B_{\text{k}})$ and $B_{\text{ext}}$ functions is weak, Fig.~\ref{fig:different_fittings_sub_Cr_Co}.
Applying the standard HH analysis without the ONE term to sub/Cr(12), while using the $M_s$, $B_{\text{k}}$, $R_{\text{AHE}}$ of the sub/Cr(10)/Co(2) and accounting for different simulated $\nabla T_{\text{Cr}}$ in sub/Cr vs sub/Cr/Co, we estimate the ONE contribution: $\xi^E_{DL} = (+48 \pm 6) \times 10^3 $\,$\Omega^{-1}$\,m$^{-1}$.
Subtracting this from sub/Cr/Co value of $\xi^E_{DL} = (+34 \pm 5) \times 10^3 $\,$\Omega^{-1}$\,m$^{-1}$, we estimate the SOT efficiency in sub/Cr/Co corrected for the ONE in Cr to be $\xi^E_{DL} = (-14 \pm 8) \times 10^3 $\,$\Omega^{-1}$\,m$^{-1}$.
This value agrees with the one obtained by fitting the high magnetic field data with eq.\,(\ref{eq:fitting_with_ONE}), Fig.~\ref{fig:different_fittings_sub_Cr_Co}.
For the reasons stated earlier, we believe that this small negative value is a more accurate estimation of the DL-SOT efficiency in Cr/Co bilayers than the numbers obtained on sub/Co/Cr.
Using this procedure, even if one cannot precisely correct for the ONE in the NM, one can judge the severity of its impact on the obtained SOT values.
We refrain from applying this approach to the sub/FM/NM sample as the thermal conductivity of the NM layer grown on the FM layer, and as a result the temperature gradient in the NM layer, can differ from the sub/NM reference sample.

In Ref.~\cite{sala_giant_2022}, sub/Co/Cr bilayers were studied by the HH measurement.
The DL-SOT efficiencies on the order of $-100 \times 10^3 $\,$\Omega^{-1}$\,m$^{-1}$ were measured.
A non-saturating $\xi_{DL}^E$ thickness dependence, similar to our Fig.~\ref{fig:xi_E_for_the_multilayer_stack}, was observed in sub/Co(2)/Cr($t_{\text{Cr}}$) and attributed to a long spin-orbital conversion length in Cr.
We have demonstrated that the ordinary Nernst effect in a combination with small $R_{\text{AHE}}$ values can give rise to large spurious SOT efficiencies, comparable to those values, and to the non-saturating $\xi_{DL}^E(t_{\text{Cr}})$ thickness dependence.
Moreover, when Gd or Tb layer was inserted between Co and Cr in Ref.~\cite{sala_giant_2022}, the $\xi^E_{DL}$ increased by 3--4 times, which was assigned to a giant orbital Hall effect in Cr and efficient orbital-to-spin conversion in 4f metals.
In view of what we have presented here and noting the lower thermal conductivity of 4f elements, which leads to higher temperature gradients, see Table~\ref{tab:material_properties} for Gd properties, we suggest that these results might need to be re-examined.
A possible control experiment to rule out the ONE contribution is to verify that the enhanced SOT is present upon the inversion of the stacking order.
Alternatively, a different measurement technique, less prone to thermal artifacts, such as magneto-optic Kerr effect~\cite{fan_quantifying_2014}, could be used to confirm the results.


\section{Conclusions}

In this work, we studied the spin-orbit torque in Cr/Co bilayers using the harmonic Hall measurement.
We performed several key auxiliary experiments: (1) reversing the layer order FM/NM vs NM/FM, (2) measuring a single NM layer, and (3) studying special multilayers designed to harbor no spin-orbit torque.
In (1), instead of the spin-orbit torque of the opposite sign and similar magnitude, we observed very different values of the same sign. 
In (2) and (3), instead of observing no SOT-like signal, we observed voltages that can be interpreted as a sizable SOT.
We demonstrated that (1), (2), and (3) can be explained by a combination of the ordinary Nernst and shunting effects.

Based on our experiments and simulations, we suggest a strategy to minimize the impact of the thermal effects in harmonic Hall measurements.
The sub/NM/FM stacking order is preferable as it significantly reduces the thermal gradient in the FM layer in comparison with the sub/FM/NM stacking.
This minimizes the ANE and ONE contributions from the ferromagnet, though increasing the ONE contribution from the nonmagnetic layer.
The latter can be estimated by conducting measurements (2) on a
single NM reference layer.
This way, even if it is not possible to precisely correct for the ONE in the NM, one can gauge the severity of its impact on the obtained SOT values.

Lastly, we estimated the ordinary Nernst coefficient in Cr and found it comparable in magnitude to the values reported for other metals.
We conclude that, despite the common belief, the ordinary Nernst effect should be carefully considered in the Harmonic Hall measurements, not only in the case of topological insulators and semimetals, but also in the case of metals.

\section*{Acknowledgements}

This work was supported by DOE Basic Energy Science Award No. DE-SC0009390.
The measurements were supported by the Army Research Office Award Number: W911NF-24-1-0152.

\section*{Data Availability}
The data supporting the findings of this article are openly available at Zenodo repository~\cite{zenodo_data_set_ONE_in_HH_measurements_2026}.

\bibliography{refs.bib}

\end{document}